\documentclass[conference]{IEEEtran}

\usepackage[T1]{fontenc}

\usepackage{soul}

\usepackage{url}
\usepackage{graphicx}
\usepackage{collcell}
\usepackage{color}
\usepackage{filecontents}
\usepackage{mathtools}
\usepackage{amsmath}
\usepackage{subcaption} 
\expandafter\def\csname ver@subfig.sty\endcsname{}
\usepackage{multirow}
\usepackage{amstext}
\usepackage{amssymb}
\usepackage[table]{xcolor}
\usepackage{tcolorbox}

\definecolor{babyblueeyes}{rgb}{0.63, 0.79, 0.95}
\definecolor{babyblue}{rgb}{0.54, 0.81, 0.94}
\definecolor{beaublue}{rgb}{0.74, 0.83, 0.9}
\begin{document}
\title{RFexpress! -- Exploiting the wireless network edge for RF-based emotion sensing}

\author{\IEEEauthorblockN{Muneeba Raja}
\IEEEauthorblockA{School of Electrical Engineering\\
Aalto University, Finland\\
Email: muneeba.raja@aalto.fi}
\and
\IEEEauthorblockN{Stephan Sigg}
\IEEEauthorblockA{School of Electrical Engineering\\
Aalto University, Finland\\
Email: stephan.sigg@aalto.fi}\\
}

\maketitle

\begin{abstract}
We present \textit{RFexpress!} the first-ever network-edge based system to recognize emotion from movement, gesture and pose via Device-Free Activity Recognition (DFAR).
With the proliferation of the IoT, also wireless access points are deployed at increasingly dense scale. 
in particular, this includes vehicular nodes (in-car WiFi or Bluetooth), office (Wlan APs, WiFi printer or projector) and private indoor domains (home WiFi mesh, Wireless media access), as well as public spaces (City/open WiFi, Cafes, shopping spaces).
Processing RF-fluctuation at such edge-devices, enables environmental perception. 
In this paper, we focus on the distinction between neutral and agitated emotional states of humans from RF-fluctuation at the wireless network edge in realistic environments. 
In particular, the system is able to detect risky driving behaviour in a vehicular setting as well as spotting angry conversations in an indoor environment.
We also study the effectiveness of edge-based DFAR emotion and activity recognition systems in real environments such as cafes, malls, outdoor and office spaces. 
We measure radio characteristics in these environments at different days and times and analyse the impact of variations in the Signal to Noise Ratio (SNR) on the accuracy of DFAR emotion and activity recognition. 
In a case study with 5 subjects, we then exploit the limits of edge-based DFAR by deriving critical SNR values under which activity and emotion recognition results are no longer reliable. 
In case studies with 8 and 5 subjects the system further could achieve recognition accuracies of 82.9\% and 64\% for vehicular and stationary wireless network edge in the wild (non-laboratory noisy environments and non-scripted, natural individual behaviour patterns).
\end{abstract}

\IEEEpeerreviewmaketitle

\section{Introduction}
Activity recognition leveraging radio frequency (RF) signal fluctuation has been explored intensively in recent research~\cite{Abdelnasser2015,Sigg,Pu:2013,wang2015review,Scholz2013DevicefreeAD}. 
These studies demonstrate that electromagnetic signals that are ubiquitously generated by cellular systems, wifi installations, Bluetooth or FM radio contain features that enable detailed human activity and gesture recognition, without binding the user with wearable sensor devices (device free activity recognition (DFAR)). 

Although recently mostly specialized equipment and software radios have been exploited, RF-based device-free recognition is best instrumented in the wireless network edge.
There, the RF-signal fluctuation observed at the wireless interface can be analyzed first-hand and processed locally.

However, most existing work has been carried out in isolated, controlled indoor environments and Line-of-Sight (LoS) scenarios, thereby ignoring many factors that affect realistic radio wave propagation such as noise, path-loss, attenuation or multipath-fading. 
For edge-based application of DFAR, it is important to investigate the system performance under 'realistic' (in contrast to 'optimal') conditions. 
DFAR basically exploits signal noise induced by movement in the proximity of a receiver. 
The accuracy of the recognition is dictated by the significance that the fingerprint of such movement leaves on the received signal.
For instance, the impact (in dB) that blocking of a signal path or constructive/destructive interference can have on the received signal is constrained in the first place by the strength of the signal itself.  
If the induced change lies below the Signal-to-Noise Ratio (SNR), the respective movement is unrecognisable for the DFAR system\footnote{Observe that typical noise resilience schemes in data communication, such as source coding and modulation, are not applicable to DFAR since the movement-induced pattern is in the traditional sense considered as noise itself}.

\begin{figure}
\includegraphics[width=\columnwidth]{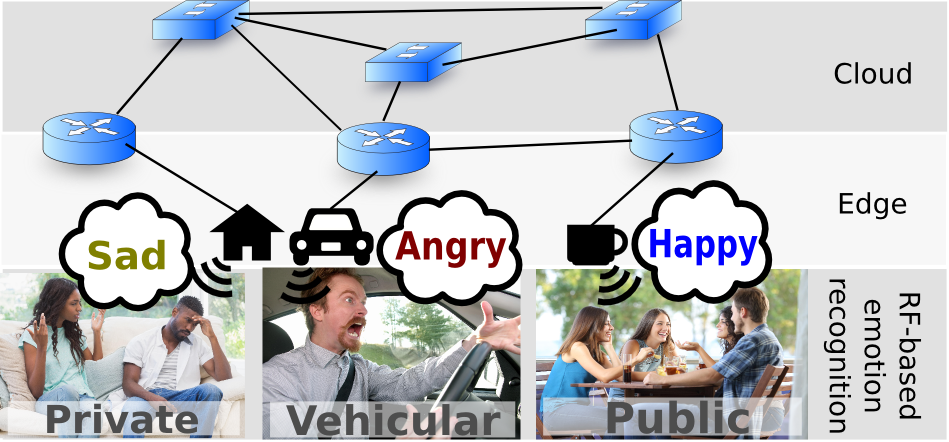}
 \caption{RFexpress! detects emotional states from RF-signals.
 }
 \label{figureTeaser}
\end{figure}
We propose \textit{RFexpress!}, an emotion recognition system that can be instrumented on standard wireless edge equipment such as WiFi or Bluetooth access points. 
As illustrated in Figure~\ref{figureTeaser}, \textit{RFexpress!} exploits signal-fluctuation patterns recognizable at the RF-interface of an edge device to distinguish emotion-indicating movement, gesture and pose. 
This information is then exploited and processed at the local edge device for the prediction of emotion and corresponding feedback. 

The recognition of emotion from DFAR systems, first suggested in~\cite{Raja}, has been exemplified also in~\cite{RFsensing_Zhao_2016} by reading pulse and respiration rate from RF-reflections for the recognition of emotional states. 
These systems require ambitious measurement equipment and a controlled environment (such as hospitals) in which disturbance through movement in the environment is prohibited.

\textit{RFexpress!}, in contrast, exploits more noisy, time-domain signal strength patterns, e.g. from RSSI fluctuation~\cite{Sigg2014}, 802.11n OFDM Channel State Information (CSI)~\cite{shi2016probabilistic} or FM radio broadcast~\cite{Sigg}. 
We investigate the impact of natural, noisy RF-environments on the classification performance of DFAR activity recognition systems like \textit{RFexpress!}, utilising features from RF time-domain signal strength which are readily available from conventional devices at the network edge.
The contributions of this paper are
\begin{enumerate}
 \item \textit{RFexpress!}: The first-ever wireless network edge-based DFAR emotion recognition system exploiting body movement, gesture and pose
 \item \textit{Wireless network edge characteristics}: A concise study of radio characteristics experienced at the wireless network edge of typical indoor and outdoor environments 
 \item \textit{Limitations}: Identification of critical SNR levels for edge-based DFAR of relevant emotion-indicating gestures (case study with 5 subjects)
 \item \textit{Vehicular edge}: Exploitation of \textit{RFexpress!} concepts for the first-ever RF-DFAR-based driver assistant system to detect risky driving behaviour (in-car case study with 8 subjects using a driving-simulation)
 \item \textit{Stationary edge}: Exploitation of \textit{RFexpress!} concepts for human motion-based emotion recognition (non-scripted case study with 5 subjects)
\end{enumerate}

To the best of our knowledge, this is the first paper to determine the impact of SNR on activity and gesture recognition. 
Furthermore, the uniqueness of this study extends to the use of gesture recognition for emotion sensing.  
We show that despite differences in body language and habit, different human emotions can be recognized.

The rest of the paper is structured as follows. Section~\ref{sectionRelatedWork} presents the research done in emotion sensing, highlighting body movements and then famous works in DFAR technology. Section~\ref{sectionArchitecture} describes the modular architecture of \textit{RFexpress!}. The study about impacts of SNR on gesture recognition is covered in Section~\ref{sectionSNR}. The real world use cases for emotion sensing using DFAR are discussed in Section~\ref{sectionExperiments}.  The conclusions and future work are listed in Section~\ref{sectionConclusion} and~\ref{sectionFuturework} respectively.

\section{Literature Review}\label{sectionRelatedWork}
In this section, we highlight previous research on (1) emotion recognition (2) activity recognition from RF signals. 
\subsection{Body Movements and Gestures for Emotion Sensing}
Human emotion recognition has been a centre of attraction for various domains particularly biology, psychology, neural networks, human computer interaction and linguistics.  
\begin{table}
\setlength{\arrayrulewidth}{0.3mm}
\setlength{\tabcolsep}{6pt}
{\rowcolors{3}{babyblueeyes}{beaublue}
\begin{tabular} { |p{1.3cm}|p{6.8cm}|} 
 \hline
\rowcolor{lightgray} Emotions & Corresponding body movements/gestures \\ [1.2ex] 
 \hline
 \rowcolor{beaublue}Anger & strong arm movements, increased knocking intensity, hands on waist, torso direction change during dancing \\ 
 \hline
 Fear& hiding body parts, contracted body movements, tensed muscles, running, backing body, hands raised and contracted with chest \\
 \hline
 Happy& open movements, clapping, extended posture, fists high up in the air  \\
 \hline
 Sad& tapping fingers, weak movements, slow walking, vertical head movements \\
 \hline
 \end{tabular}}
 \captionof{table}{Set of gestures, body movements and postures mapping to particular emotions used in previous research.}
 \label{bodymovements}
\end{table}
The most well known emotion sensing methods exploit modalities such as facial expression~\cite{Baggio20122121,Martino2011,Schmid2011}, speech~\cite{Ayadi2011,Schroeder2006}, text~\cite{Paltoglou2012,Wang2012}, physiological signals (heart rate, breathing rate)~\cite{Chanel2011}, input devices (keyboard, mouse)~\cite{Lv2008,Kolakowska2013}, body movements and gestures~\cite{Burgoon2010,Chen2011} or combination of various features~\cite{Busso2004,Soleymani2012}.  
Widely considered emotions are sadness, happiness, anger, fear, surprise and disgust~\cite{Karg2013}. 
Most common modalities are facial expressions (more than 95\%~\cite{GelderRuudHortensius2009}).
However they come with challenges like privacy intrusion, required high image quality and computationally intensive algorithms~\cite{Sandbach2012}. 
On the other hand, textual input can be deceptive due to contextual dependency, speech recognition suffers from noisy speech signals and dependency on region and language~\cite{Koolagudi2012}. 
Physiological signals are highly reliable but require extreme human involvement due to body worn devices and restricted movement for accurate detection~\cite{Wioleta2013}. 

Body movements are an underestimated modality for detecting human emotions~\cite{Nele2012}, gaining popularity in recent studies, due to low cost and increasingly reliable body sensing technologies. 
Body expressions outperform facial expressions when emotions are to be detected from larger distances~\cite{GelderRuudHortensius2009}, for discriminating strong emotions (positive vs. negative)~\cite{Aviezer2012} and for emotions which cant be socially altered (e.g. fear and deception)~\cite{GelderRuudHortensius2009}.
Unlike facial expressions, body movements are a not much affected by culture and gender and are better modality for crowd sensing. 
Table~\ref{bodymovements} summarises some gestures and their link to emotion (cf.~\cite{Karg2013,Karg2010,Gunes2007,Nicolaou2011,Zhang2011}).

\subsection{Body Movements and Gesture Recognition from RF}
Activity and gesture recognition from RF is an active research topic~\cite{Qi2012}.
The fluctuation in the multipath propagation observed at a receive antenna indicate human movements, gestures or environmental conditions~\cite{Sigg2014,Pu2013,Adib2013}.  
DFAR exploits existing infrastrucure and overcomes device boundedness and limited range of traditional sensing~\cite{Wang2015}.  
Adib et al.~\cite{Adib2013} use WiFi signals and MIMO interference to track the body movements and gestures behind the walls without requiring any on-body sensors. 
Similarly, Pu et al.~\cite{Pu2013} perform hand gesture and human movement recognition via analysis of micro Doppler fluctuation and signal distortions. 
Wang el al.~\cite{Wang2014} detect inplace, walking movements and daily home activities by taking CSI signals as location activity profiles. 
Up to eight gestures (flick, pull, push, punch, lever, zoom in, zoom out, double flick) can be detected with a low cost wireless signal based method~\cite{Kellogg2014}. 
In addition, recently presented systems utilise fresnel effects for breathing detection~\cite{wang2016human} or phase-variation for gait estimation~\cite{wang2016gait}.
These systems, however, are hardly feasible in realistic environments due to their fragile requirements for the setup (exact distance and orientation between subject and receiver).

The recognition capabilities with off-the-shelf equipment in everyday installations, parasitically utilising environmental signals from pre-installed systems (e.g. FM-radio, WiFi, 4G) are restricted to less fine-grained movement but still allow decent recognition of activity classes. The recognition of walking speed and hand gestures by monitoring the signal strength of environmental WiFi routers from standard mobile phones has been demonstrated in~\cite{Sigg2014,Abdelnasser2015}. 

Body movements map to human intentions, attention~\cite{Shi2014} and emotions~\cite{Pervasive_Jaggarwal_2012,Emotion_Castellano_2008}, DFAR can, by detecting gestures or body movements, indicate respective emotions. 
However, in order to practically incorporate body movement detection with emotion sensing, we need to understand how wave propagation losses in real environments (as opposed to lab setups) may limit the use of RF for activity detection and in turn emotion sensing. 

\section{\textit{RFexpress!} architecture}\label{sectionArchitecture}
\begin{figure}
\centering
  \includegraphics[width=\columnwidth]{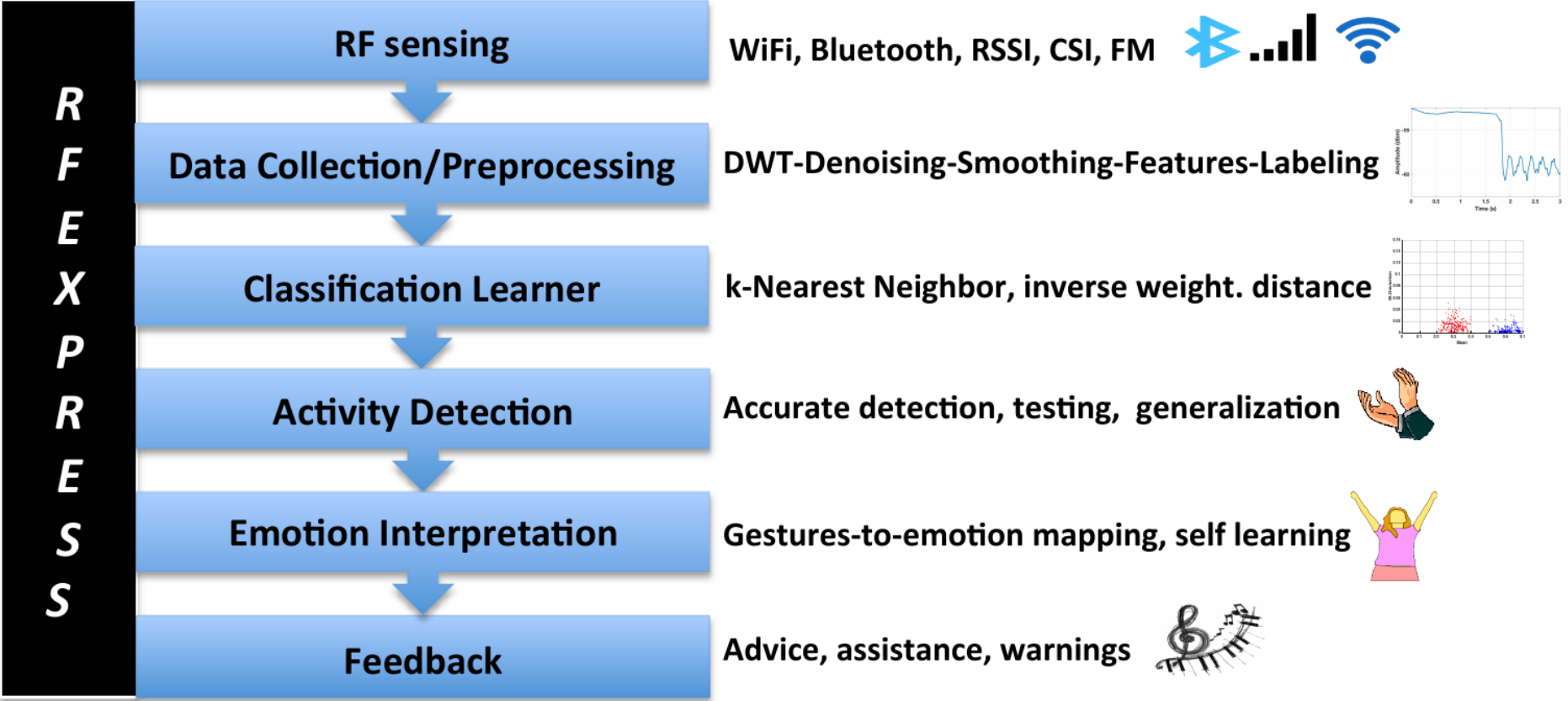}
   \caption{Modular architecture of \textit{RFexpress!}. }
 \label{RFexpress}
\end{figure}

\textit{RFexpress!} follows a modular structure, similar, for instance, to related DFAR systems such as~\cite{Abdelnasser2015}. 
We will briefly overview the respective modules and leave the technical details for the discussion in the experimental section~\ref{sectionExperiments}. 
The system is depicted in Figure~\ref{RFexpress}. 
The RF stimuli of environmental movement and activities are captured by the RF sensing module, which directly interfaces the wireless channel extracting. 
We exploit RSSI information in our measurements in section~\ref{sectionSNR} but other possible data includes Channel-State-Information (CSI), Bluetooth, or FM-radio. 
This continuous stream of data is then denoised (we utilise discrete wavelet transformation in our implementation) and smoothened.
The (non-overlapping) windowing, feature computation and data labelling is performed on processed data and features are forwarded to the classification learner. 
The classification learner is responsible for the model training, which includes data partitioning, training and achieving a model for classifying gestures.
In our implementation, we implemented a k-nearest neighbour classifier but other classification modules can be exploited interchangeably. 
The trained classification module is, as a next step, exploited to predict meaningful activity classes. 
These classes (movement, gestures and activities) are then mapped to emotional states (cf. section~\ref{sectionRelatedWork}).
In our case, in section~\ref{sectionExperiments}, we distinguish between angry and normal emotion states. 
This emotion-information is then used for the the generation of application-dependent feedback such as, for instance, to prevent risky driving behaviour (see section~\ref{sectionExperiments}).
We utilise and discuss these modules in the following sections.

\section{Impact of SNR on gesture recognition}\label{sectionSNR}
\begin{figure}
 \centering
  \includegraphics[width=\columnwidth]{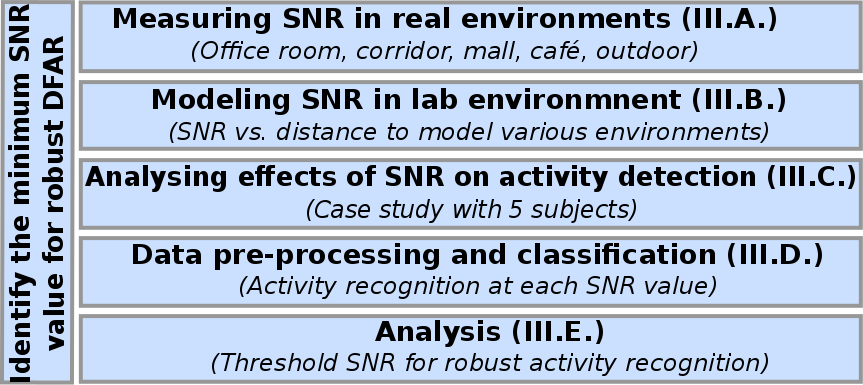}
  \caption{Vital steps performed to achieve the critical SNR values and understand the impact on activity recognition. }
 \label{snrStudy}
\end{figure}

We determine the effectiveness and accuracy of DFAR systems in real environments in comparison to controlled environments. 
To achieve this we first measure the radio characteristics of classical real environments where the signal-to-noise ratio (SNR, $S=P_S/P_N$) is the primary parameter~\cite{Kieser2005}.
SNR describes the upper boundary how clear any gesture or activity can be observed from RF because any detectable fluctuation is necessarily above noise.
Then, we model these SNR values for more detailed study on the DFAR performance, and perform case studies with 5 subjects to identify critical SNR values for robust activity recognition (cf. Figure~\ref{snrStudy}).

\subsection{Measuring SNR Values in Real Environments}
We have considered 5 different real environments for our SNR study, including office room, corridor of a university building, mall, cafe and outdoor. 
For each environment, we measure the signal strength and noise between a wifi router\footnote{Cisco 802.11 g Linksys WRT54G/GL/GS, Tx frequency: 2.43GHz, TX Power=251mW, rate=36Mbps} and a laptop (receiver) at different distances. 
The experiment for each environment is repeated 5 times on different days and times of the day. 
For each measurement, packets are traced for 60 seconds with Wireshark to obtain signal strength and noise of each packet. 
We calculate SNR by taking signal and noise power level differences, and then take an average of all SNRs: $SNR(db)=P_S(dbm)-P_N(dbm)$.

\begin{table}
\setlength{\arrayrulewidth}{0.3mm}
\setlength{\tabcolsep}{6pt}
\begin{center}
{\rowcolors{3}{babyblueeyes}{beaublue}
\begin{tabular} { |p{2cm}|p{0.85cm}|p{0.85cm}|p{0.85cm}|p{0.85cm}| p{0.85cm}|} \hline
\rowcolor{lightgray}\multicolumn{6}{|c|}{SNR Values (db)} \\
 \hline
Environment & 0m & 8m & 17m & 25m & 30m \\ [1.2ex] 
 \hline
 Cafe & 77.3 & 42  & 20.6 & 5&\\ 
 \hline
 outdoor& 60.7 & 45.1 & 44.9 & 36.7&37.3\\
 \hline
 office& 74.3 & 52.1 & 41.6 & 30.6 & \\
 \hline
 building corridor& 76.09 & 49 & 56.4 & 25 & 5 \\
 \hline
 mall& 71.3 & 46.49 & 40.2 & 36.51& 25.9\\ 
 \hline
 \end{tabular}}
 \captionof{table}{Average SNR observed in all environments for various distances between the transmitting WiFi router and the laptop}
 \label{snrTable}
 \end{center}
\end{table}

Table \ref{snrTable} shows the SNR with changing distances in different environments. 
For all the environments the SNR is  between 60db and 80db when the receiver and transmitter are placed next to each other (0m apart). 
Naturally, the SNR decreases with increasing distance. 
The corridor and mall (which also has a  corridor layout) show high SNR even at larger distances. 
The outdoor environment suffers from an initial low SNR but stays steady with increasing distance. 
Cafe environment shows a tremendous drop in SNR with increasing distance. 
While these general trends are expected, the measurements give us a better understanding what recognition performance to expect for DFAR systems under various conditions as detailed below.

\subsection{Modelling SNR values in Lab Environment}
\label{snrSection}
\begin{figure}
 \centering
  \includegraphics[width=\columnwidth]{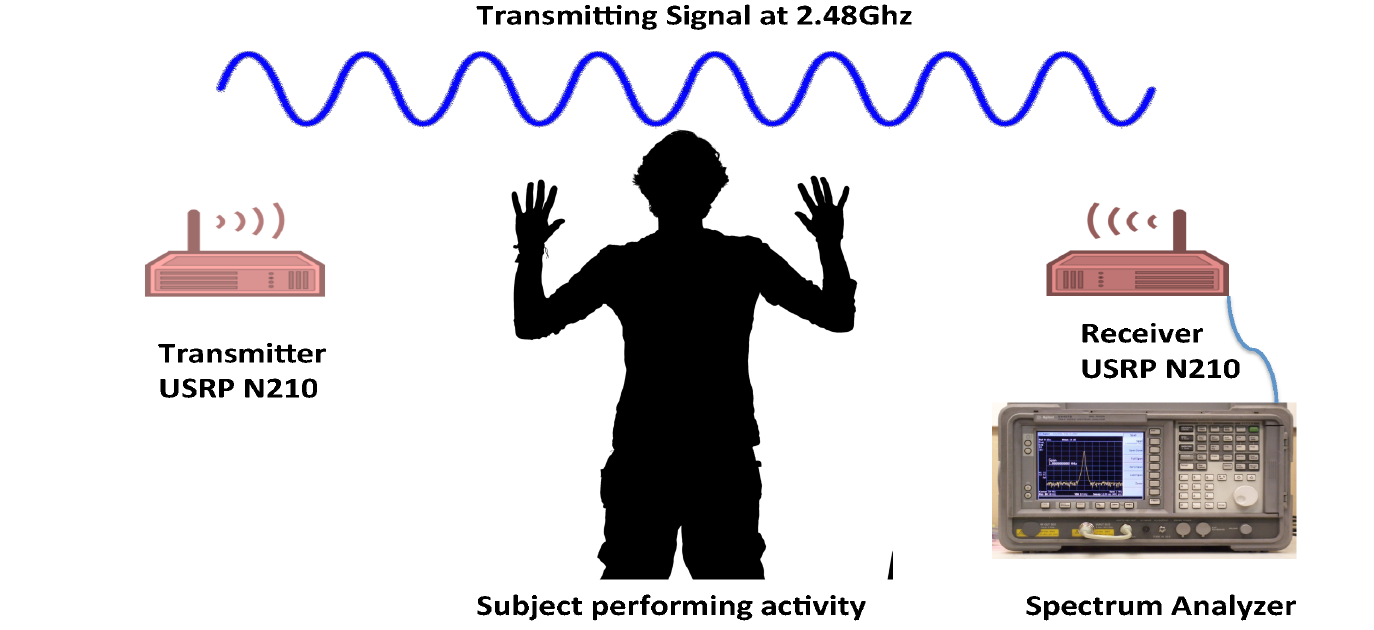}
  \caption{Experimental set up for gestures activity detection of active system. }
 \label{experimentSetup}
\end{figure}

The SNR vs. distance values obtained from different environments provided us the range of SNR values to model in our lab environment. 
We set up USRP (N210) SDR transmit and receive devices (Figure~\ref{experimentSetup}) with SBX daughterboards (400 MHz - 4.4 GHz) and omni-directional antennas. 

At the location of the receive USRP we utilise a spectrum analyser (Rohde \& Schwarz 20Hz-3.5Ghz, FSEA) to measure the power levels at the receiver end. 
The transmitter and receiver are kept 2m apart. 
All characteristics in the lab like, furniture, equipment and people are constant. 
In order to calculate SNR from USRPs, we first measure the ambient noise by detecting power (dbm) at the receiver while the transmitter is off. 
The ambient noise power value is negligible, and it is even lower than the spectrum analyser's own thermal noise and remains constant throughout our experimentation. 
Therefore, in order to generate the desired SNRs, we introduce the noise at the transmitter (in software via Gnuradio) and measure the value from the spectrum analyser.  
In order to get the signal strength, we measure the power level at the receiver when during transmission and subtract the noise value that we obtained before. 
The transmitter USRP transmits a sine wave of frequency 100kHz with a sampling rate of 1MHz and centre frequency of 2.48GHz. 
We can then tune power and noise values at the transmitter to obtain the desired SNR values at the receiver. 
SNR is derived straightforward:
\[dBm=10Log(Power/1mW) \]  
where \textit{Power} is the combined value of $P_S+P_N$ in dbm measured at the receiver, we convert it into mW to get $P_S$ and $P_N$ separately.
\begin{eqnarray}
dBm/10=log(Power/1mW) \nonumber \\
Power(mW)=Log^{-1}(dBm/10)\nonumber \\
Signal(mW)=Power(mW)-Noise(mW) \nonumber \\
SNR=10log(signal(mW)/Noise(mW))  \nonumber
\end{eqnarray}

\subsection{Modelling Real Environments and Analysing Effects of SNR on Activity Detection }
Activity detection at 6 different modelled SNR values (59dB, 42dB, 22dB, 12dB, 2dB and 0dB) has been tested with 5 subjects (3 males, 2 females). 
For each SNR, the person repeats the experiment 5 times with a system trained on the following three activities.
As our ultimate goal is to bridge a link between gesture detection and emotion sensing, we choose gestures for this experiment which map to certain emotional states~\footnote{We remark that the purpose of this study was not on the emotion recognition itself but on the impact of the SNR on the recognition accuracy. More decent analysis is required to comprehensively recognize particular emotion from body movement}: 
(1 -- neutral) hands down when a person is standing in between receiver and transmitter at 0 degree orientation, (2 -- fear/shock) hands raised up to shoulder level, and (3) clapping while hands are down. 
For case (2) and (3), participants were instructed that they should show their upper body reaction when facing an emergency situation in which a bear is approaching them (shock/fear) and when applauding in happy state in a musical evening or watching favourite match.
 \begin{figure*}
\centering
\begin{subfigure}{.33\textwidth}
  \includegraphics[trim=50 20 300 0,width=.7\linewidth,height=3.5cm]{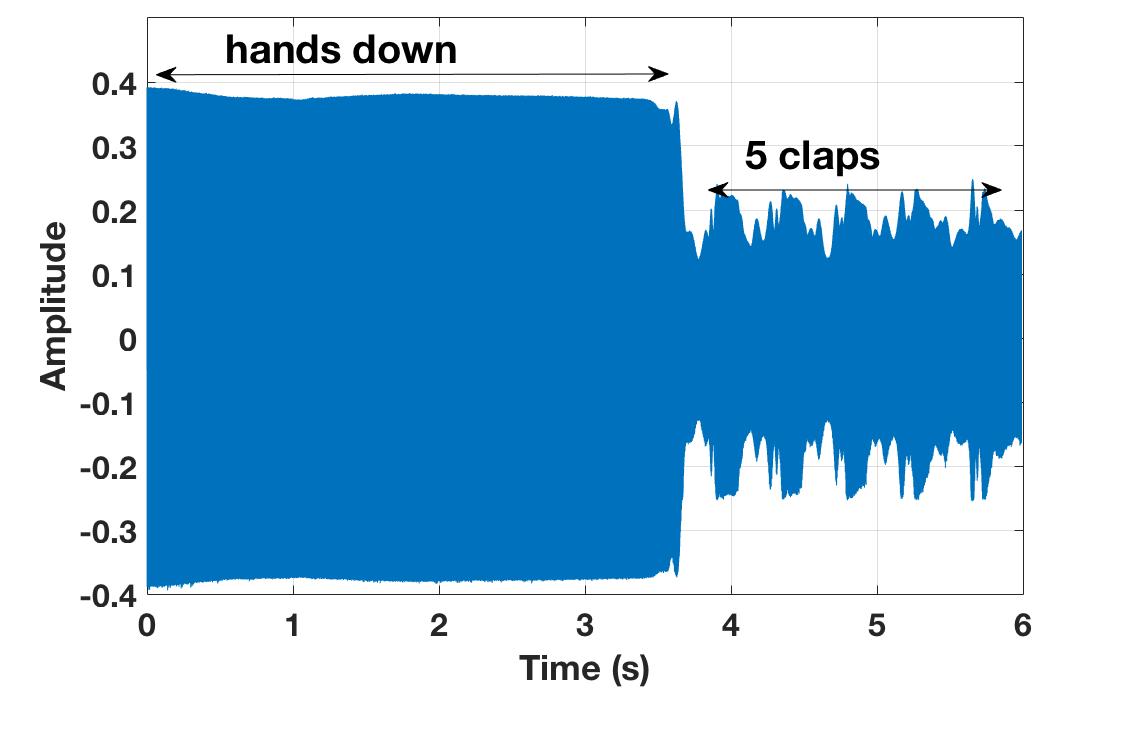}
  \caption{clipping.}
  \label{fig:cSignal}
\end{subfigure}%
\begin{subfigure}{.33\textwidth}
  \includegraphics[trim=50 0 300 35,width=.7\linewidth,height=3.5cm]{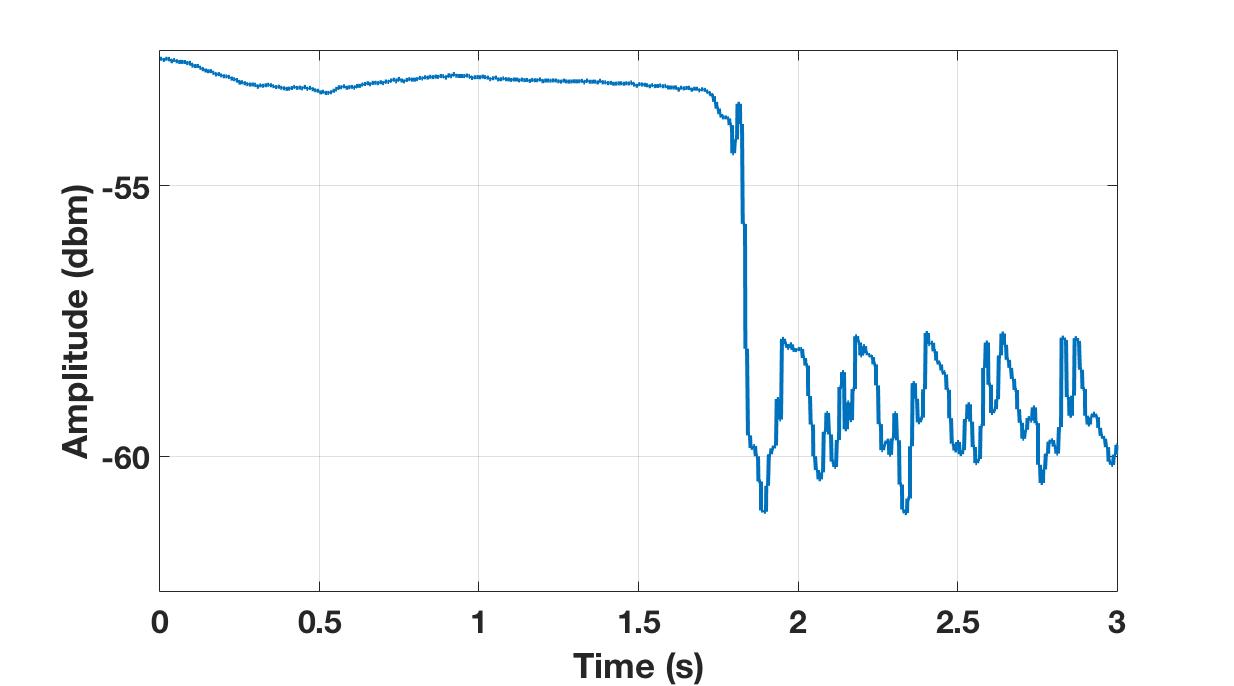}
  \caption{DWT+denoising.}
  \label{fig:pSignal}
\end{subfigure}
\begin{subfigure}{.33\textwidth}
  \includegraphics[trim=50 0 300 35,width=.7\linewidth,height=3.5cm]{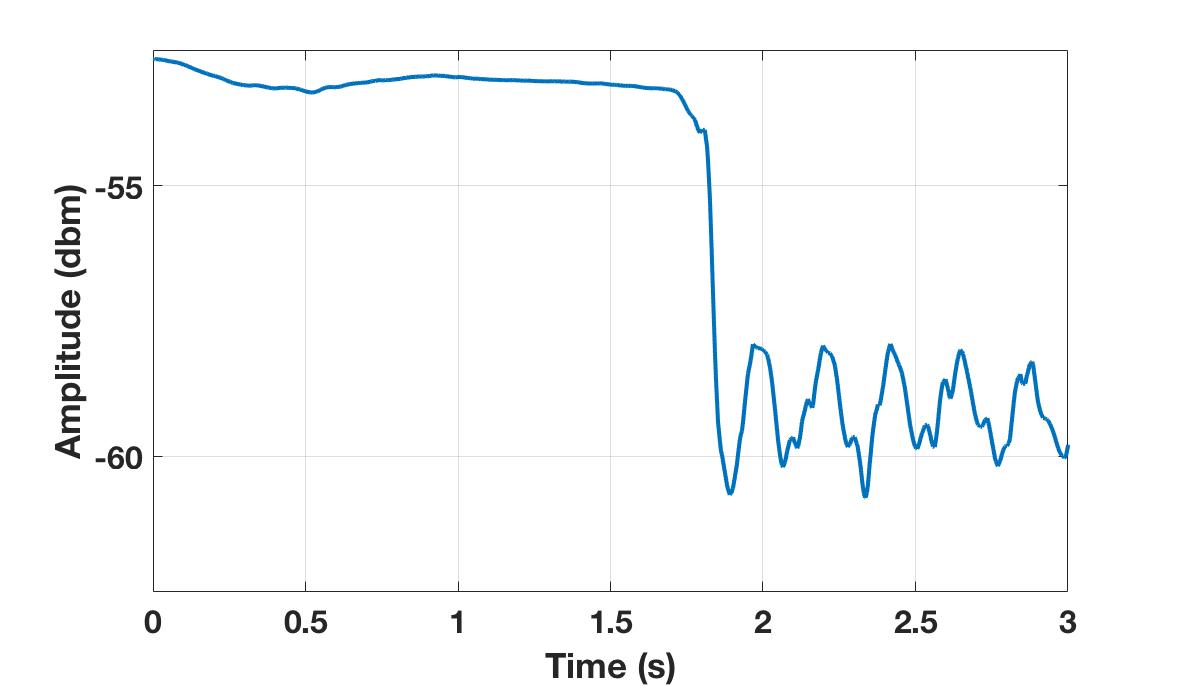}
  \caption{smoothing.}
  \label{fig:sSignal}
\end{subfigure}
\caption{Series of steps performed for signal processing in (L-R) order; clipping of the original signal, DWT(level 13, Haar function) and denoising (Stein's Unbiased Risk method)~\cite{Abdelnasser2015} and smoothing with moving average filter.}
\label{dataprocess}
\end{figure*}

\begin{table} 
 \setlength{\arrayrulewidth}{0.3mm}
\setlength{\tabcolsep}{6pt}
\begin{center}
{\rowcolors{2}{babyblueeyes}{beaublue}

\begin{tabular} { |p{1cm}|p{1.4cm}|p{2.4cm}|p{2cm}|p{2.2cm}| } \hline
\renewcommand{\arraystretch}{2.5}
SNR Value (db) & Hands up/down, d=2m (\%) & Hands up/down/claps, d=2m (\%) & Hands up/down, d=4m (\%) \\ 
 \hline
 59 & 90 & 83 & 84\\ 
 \hline
 42 & 88 & 81 & 83\\
 \hline
 22 & 79 & 68 & 69 \\
 \hline
 12 & 65 & 60 & 60 \\
 \hline
 2 & 53.5 & 47 & 49\\ 
 \hline
 0 & 52 & 41 & 42 \\
 \hline
 \end{tabular}}
 \captionof{table}{Impact of SNR on accuracy of gesture/activity detection. The 2nd column shows the classification accuracy of two activities, hands down and up when distance bw receiver/transmitter is 2m. 3rd column shows the classification accuracy with three activities; hand down, hand up, clapping. The 4th column show the accuracy of 2 activities when distance bw receiver/transmitter is 4m. The accuracy with increasing distance is not much affected because the SNR remains considerably high at 2m.}
 \label{accTable}
 \end{center}
\end{table}

\subsection{Data Pre-processing and classification} 
We preprocess the collected raw data in order to to get rid of the noise, extract detailed edges and obtain accurate results. 
The signal denoising technique in our study is inspired by Abdelnasser et al.~\cite{Abdelnasser2015} and uses Discrete Wavelet Transform (DWT) which provides both time and frequency representation of a signal for fine grained multi scale analysis. 
Figure~\ref{dataprocess} illustrates the data during processing.
Afterwards, we calculate statistical and frequency-domain features (mean, standard deviation, entropy, zero crossing and average derivative) to distinguish the activities from the data.  
The most indicative features were mean and standard deviation. 
We chose the (non-overlapping) window size to be 100.000 samples with 1MHz sampling frequency or 10 windows/second for the feature computation. Table~\ref{accTable} shows the classification accuracies achieved for target activities with respect to the SNR values.

\begin{figure*}
\centering
\includegraphics[width=\textwidth]{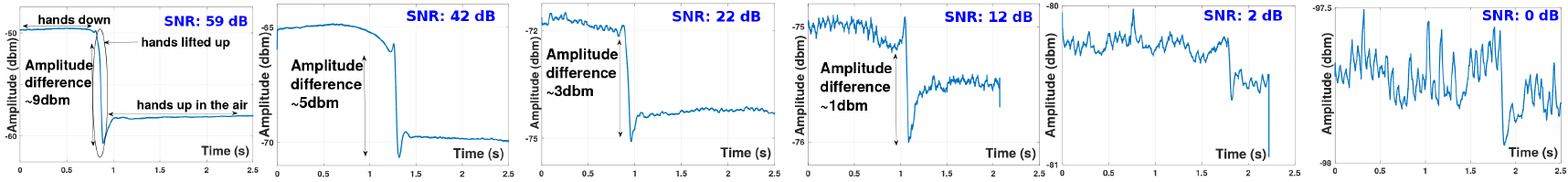}
\caption{The impact of SNR on gesture recognition. Each graph represents the processed gesture signals with various SNR values. Activities are hands down and hands up.}
\label{test2}
\end{figure*}

\begin{figure*}
\centering
\begin{subfigure}{.33\columnwidth}
  \includegraphics[width=\linewidth]{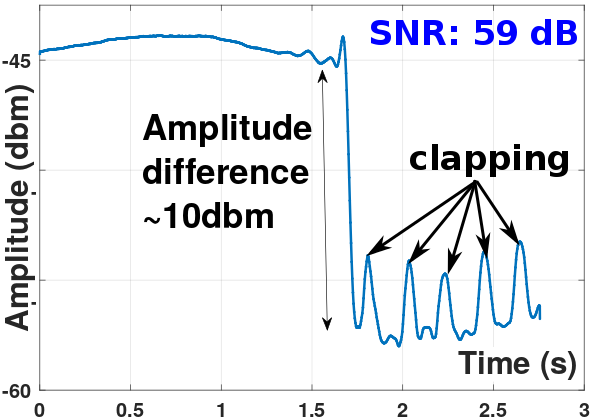}
  \label{fig:sub7}
\end{subfigure}%
\begin{subfigure}{.33\columnwidth}
  \includegraphics[width=\linewidth]{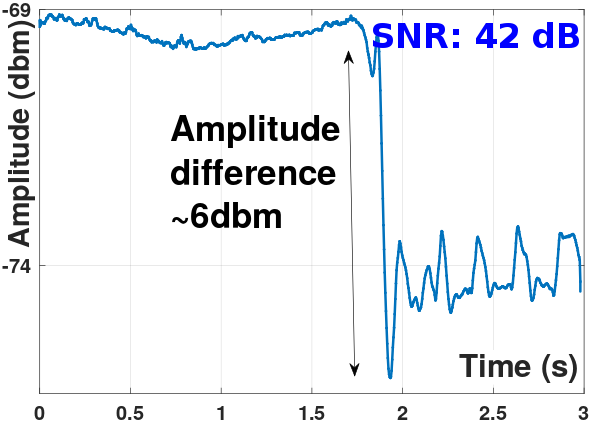}
  \label{fig:sub8}
\end{subfigure}
\begin{subfigure}{.33\columnwidth}
  \includegraphics[width=\linewidth]{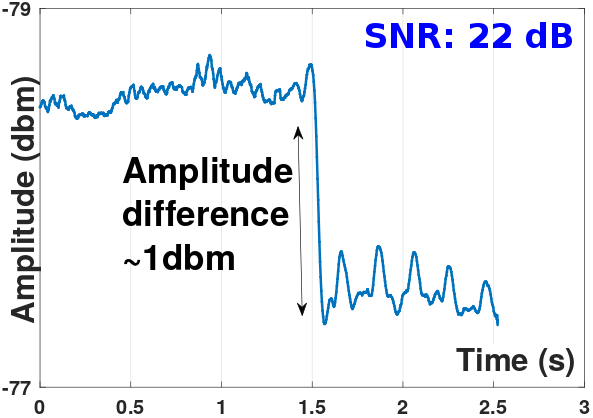}
  \label{fig:sub9}
\end{subfigure}
\begin{subfigure}{.33\columnwidth}
  \includegraphics[width=\linewidth]{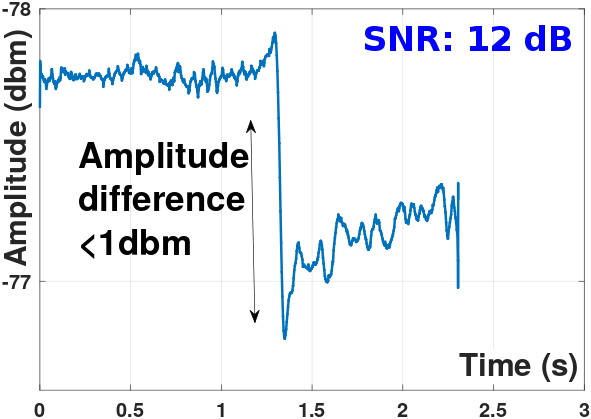}
  \label{fig:sub10}
\end{subfigure}%
\begin{subfigure}{.33\columnwidth}
  \includegraphics[width=\linewidth]{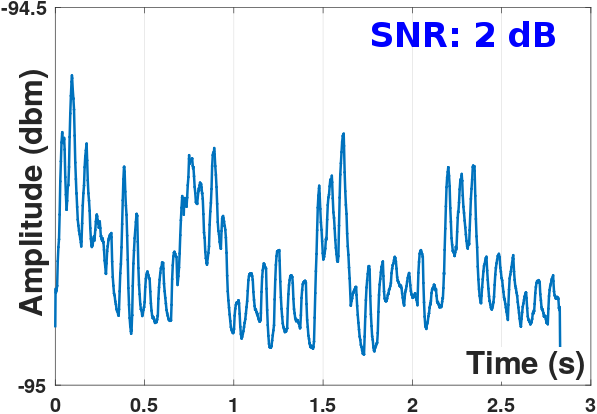}
  \label{fig:sub11}
\end{subfigure}
\begin{subfigure}{.33\columnwidth}
  \includegraphics[width=\linewidth]{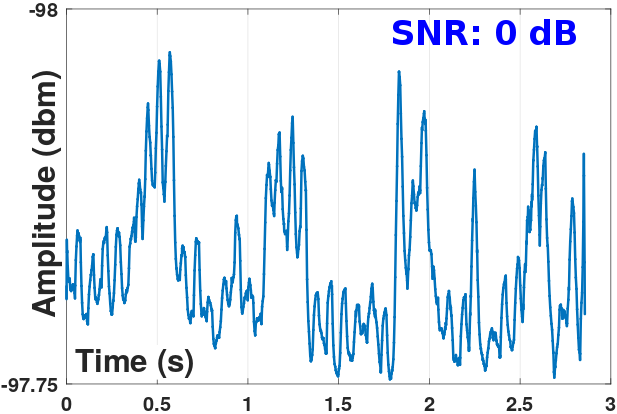}
  \label{fig:sub12}
\end{subfigure}
\caption{The impact of SNR on gesture recognition. Each graph shows the gesture recognition at different SNR value. The gestures are hands down vs. clapping.
}
\label{test3}
\end{figure*}

\subsection{Analysis}
Its obvious from the table that higher SNR foster better accuracy. 
However, this also depends on the complexity of the activity. 
For activities like hands up and down, the accuracy is reasonably high even at 22dB. 
For activities like clapping, with less significant footprint on the signal, though, the accuracy deteriorates significantly, even at high SNR. 
Increasing the distance between transmitter and receiver, however, did not have a noticeable difference in accuracy for hands up and down (cf. table~\ref{accTable})
SNR values of about 30dB and higher can be considered for robust activity recognition of different activities. 
SNR values of 20dB and below can lead to erroneous results and at lower SNRs activity can not be detected. 
The effect of SNR on gesture recognition is further visible from Figure~\ref{test2} and~\ref{test3}. 
The processed signal at 59db, 42db and 22db for hands down and up can be clearly spotted. 
Below 22db the distinction of the signal becomes challenging even in the processed signals. 
Furthermore, since the relative variance in the amplitude also decreases with decreasing SNR, Figure~\ref{test3} (hands down vs. claps) shows that, while at SNR 59db and 42db, we can even count the number of claps in processed signals, this becomes quickly challenging for lower SNR levels.

\section{Real World Applications for Emotion Recognition }\label{sectionExperiments}
We exploit \textit{RFexpress!} concepts for the distinction between different emotional states in two realistic cases: (1) Detection of risky, agitated driving behaviour and (2) detection of angry argument in an indoor setting. 
In both scenarios, WiFi installations are common (WiFi or Bluetooth in cars; WiFi access points in indoor environments) and suggest the application of DFAR. 
Road rage is a serious problem and according to the national Highway Traffic Safety Administration\footnote{http://www.nhtsa.gov/Aggressive}, aggressive driving is responsible for more than 66\% of all car crashes in the US.
Likewise, in an industrial, professional context, negatively emotional agitated states, such as anger or even rage seriously worsen the ability to successfully negotiate a maximum economic bargain~\cite{fabiansson2012effects}. 
Detecting and mediating repeated negatively emotionally aroused behaviour can therefore maximise the economic outcome.  

\subsection{Scenario 1: Detection of risky driving behaviour}
\label{carSection}
\begin{figure}
\begin{subfigure}{.337\columnwidth}
\includegraphics[width=0.9\linewidth]{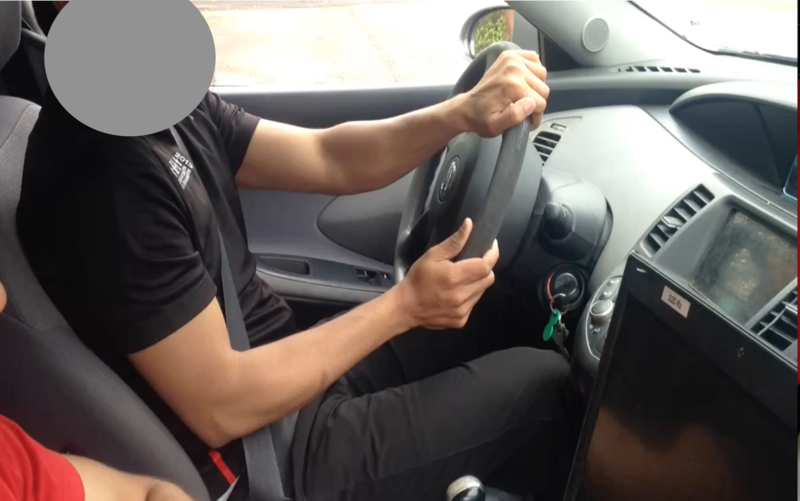}
 \caption{neutral 1 }
\end{subfigure}%
\begin{subfigure}{.337\columnwidth}
\includegraphics[width=0.9\linewidth]{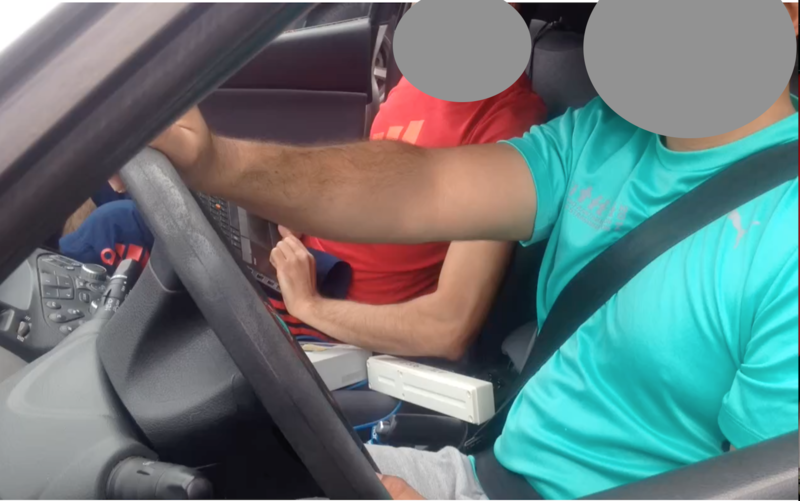}
\caption{neutral 2}
\end{subfigure}%
\begin{subfigure}{.337\columnwidth}
\includegraphics[width=0.9\linewidth]{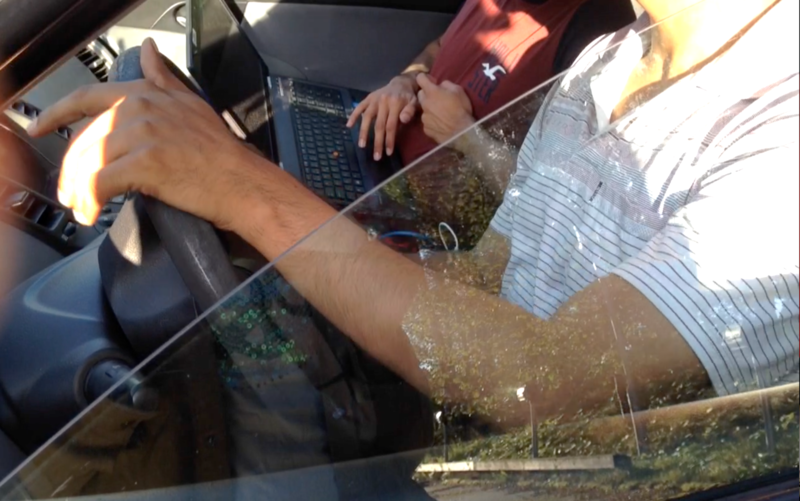}
\caption{neutral 3}
\end{subfigure}%

\begin{subfigure}{.337\columnwidth}
\includegraphics[width=0.9\linewidth]{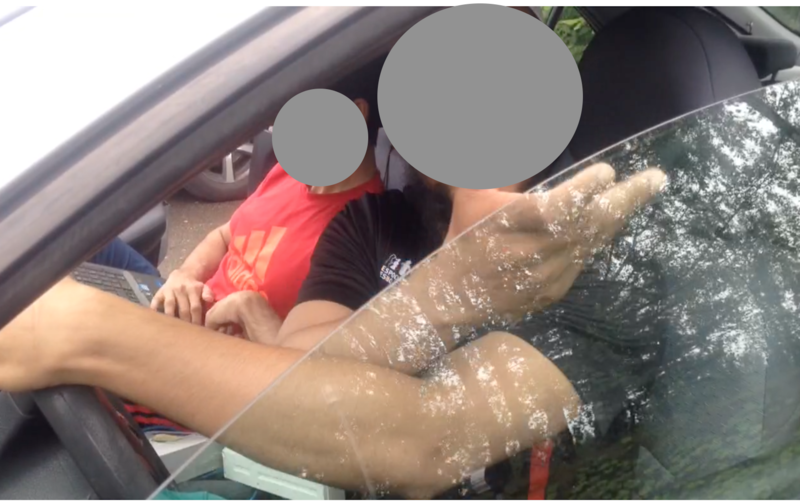}
\caption{angry 1}
\end{subfigure}%
\begin{subfigure}{.337\columnwidth}
\includegraphics[width=0.9\linewidth]{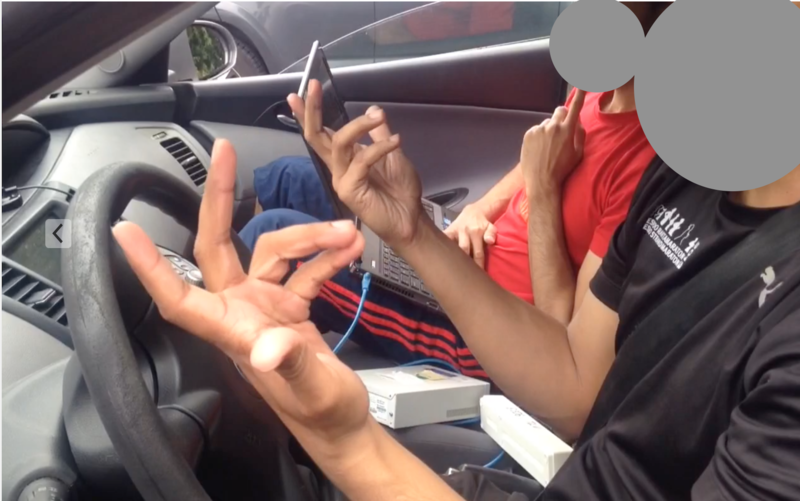}
\caption{angry 2}
\end{subfigure}%
\begin{subfigure}{.337\columnwidth}
\includegraphics[width=0.9\linewidth]{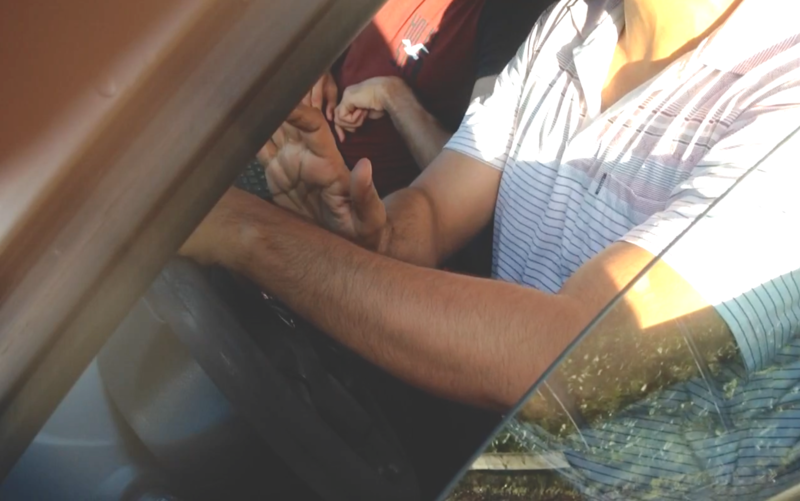}
\caption{angry 3}
\end{subfigure}%
\caption{Images captured during the driving car experiments. The top row indicates neutral driving styles of subjects. While the bottom row indicates the gestures like pointing fingers, honking horns and showing anger
}
\label{carImages}
\end{figure}

Based on our experiments and results on different SNR levels that we present in the previous section, we venture forward to evaluate the effectiveness of DFAR systems in real world scenarios that may suffer from SNR degradation. We propose the less intrusive, cheap and pervasive DFAR in a car environment, to detect strong negative emotions. 
Many cars nowadays feature in-car Bluetooth or Wifi installations, that could be readily exploited for such DFAR emotion recognition. 
Body movements and gestures are the modality or stimuli for mapping emotions in our case. 
In particular, we distinguish normal driving behaviour and driving in an angry state of the driver in her vehicle. 
\subsubsection{Experiment}
To distinguish between normal driving behaviour and angry or rage driving behaviour using RF signals, we performed a driving experiment with 8 subjects (5 males and 3 females), aged 18 to 40 and belonging to 4 different countries (German, Iran, Pakistan and Vietnam). 
The technical details of the system utilized are identical to the system described for the SNR study above. 
Also the features utilised (mean and standard deviation) and the (non-overlapping) window size of 100.000 samples with 1MHz sampling rate have been kept identical.
Each subject was assigned 30 minutes for performing the experiment and before the start of the measurements, they were given demonstration and briefing, were allowed to adjust the car settings according to their personal driving attire and were made familiar with the simulator video. 
We used a driving simulator video~\footnote{https://www.youtube.com/watch?v=lyVctz5BAro} during the course of the experiment. 
It was played on a tablet and fixed on the windscreen according to driver's preferred position. 
For normal driving, the subjects were asked to drive through the simulator video while performing normal driving, the way they usually do (steering the wheel, changing gears, checking the rear mirrors). 
For the angry driving case, we created a cover story the subjects should consider while following the same driving simulator video, acting and responding the way they would do while actually driving. 
The cover story is designed following previous research~\cite{Leng2007,DePasquale20011,Grimm2007} on emotion elicitation in which emotions are induced using texts, pictures, videos and situations for driving and other scenarios. Our cover story is:
\vspace{.3cm}

\noindent\fbox{\begin{minipage}{.95\columnwidth}
                \textit{``You have a flight to catch in an hour, and you have maximum 15 minutes to reach the airport. You have an extremely important business meeting to attend in another country and you can not afford to miss the flight. You leave home in a very frustrated mood and get stuck in the traffic. The driver in the parallel car throws his cup of coffee outside which hits your wind screen~\cite{DePasquale20011} and then overtakes you without giving an indicator. Now every crossing pedestrian or small mistake by other drivers gets you in a furious state and you express anger throughout your drive.
You also have a companion on front seat who is commenting on your driving attitude and you get into an argument with him while driving." }
               \end{minipage}
}\vspace{.3cm}

\subsubsection{Data Collection} 
\begin{figure}
 \centering 
  \includegraphics[width=0.5\textwidth]{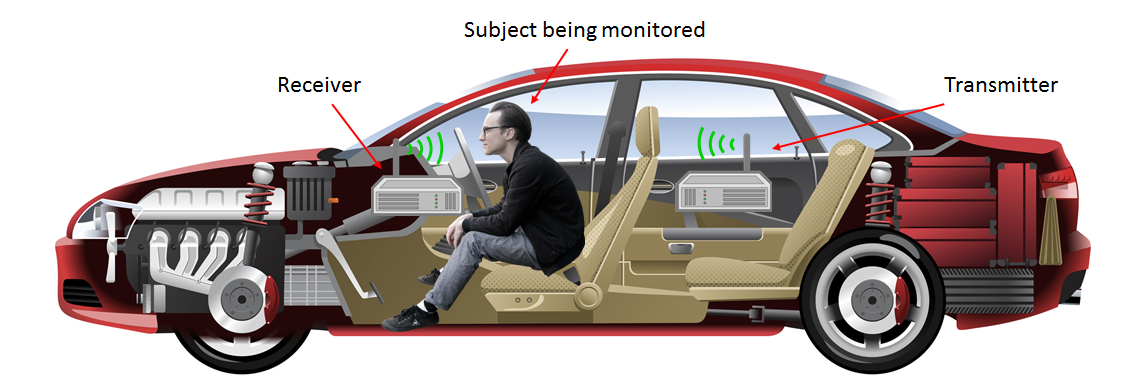}
  \caption{Experiment set up for gestures activity detection and emotion recognition in car. }
 \label{RFexperiment}
\end{figure}

We collect the data for gesture movements by setting up an active recognition system as used for SNR experiments. 
The devices set up is such that the receive antenna is fixed on the car's cockpit and the transmit antenna is placed on the headrest of the back passenger seat.  
Neutral and angry driving tests are performed 4 times for each subject. 
The images captured during the experiments are shown in Figure~\ref{carImages} and the set-up can be seen in Figure~\ref{RFexperiment}.

We use k-Nearest Neighbour (k-NN) classifier for our classification. 
We set the number of neighbours to k=6 and chose inverse weighted distance for nearest neighbour computation.  
We train the models for individual data as well as aggregated data from all subjects. 

\begin{figure*}
\centering
\begin{subfigure}{.45\textwidth}
  \centering
  \includegraphics[trim=60 235 0 255,width=0.98\linewidth,height=5.5cm]{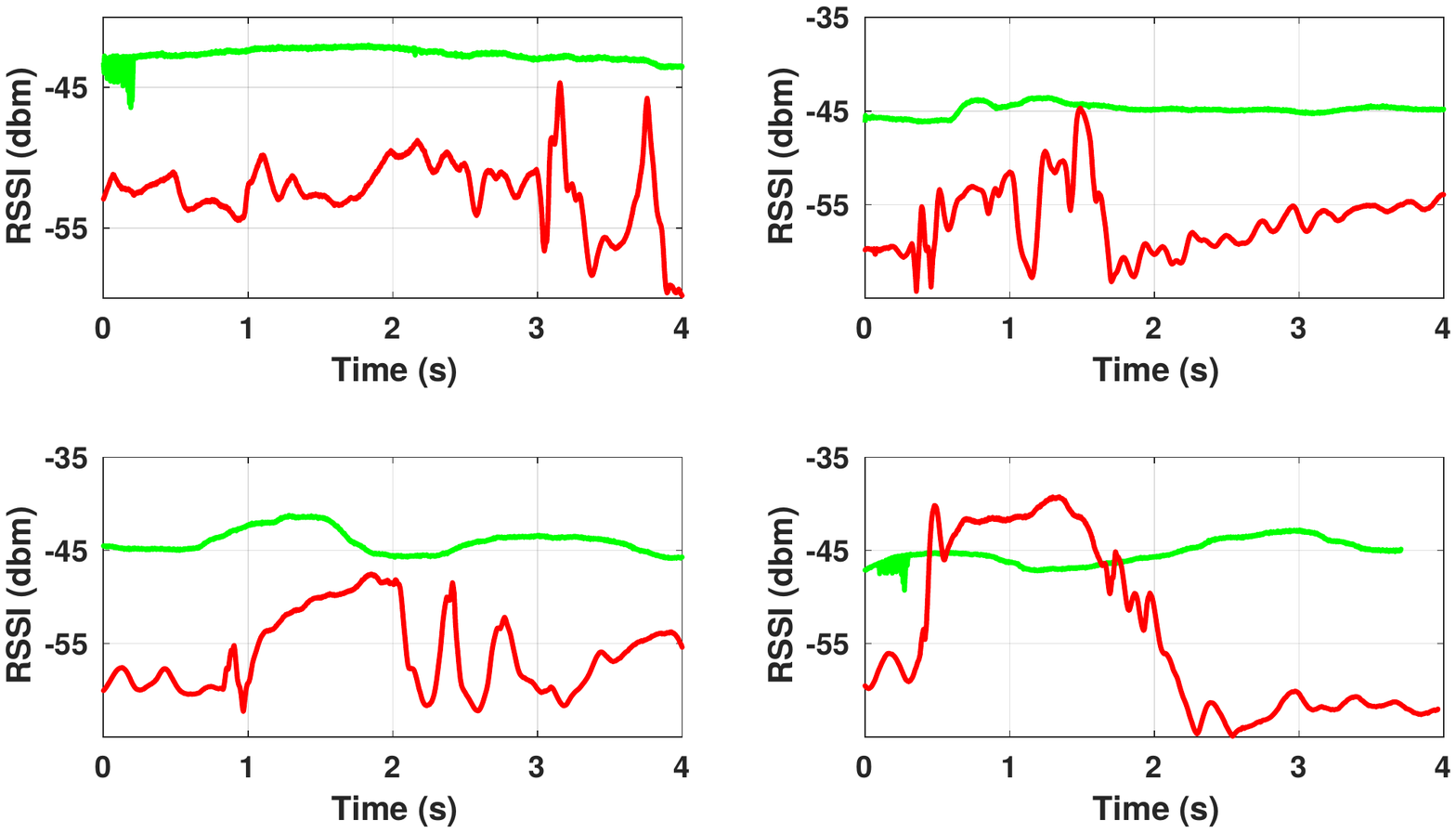}
   \caption{Neutral vs. Angry graphs of subject1 }
\end{subfigure}
\begin{subfigure}{.45\textwidth}
  \centering
  \includegraphics[trim=0 235 0 255,width=1.16\linewidth,height=5.5cm]{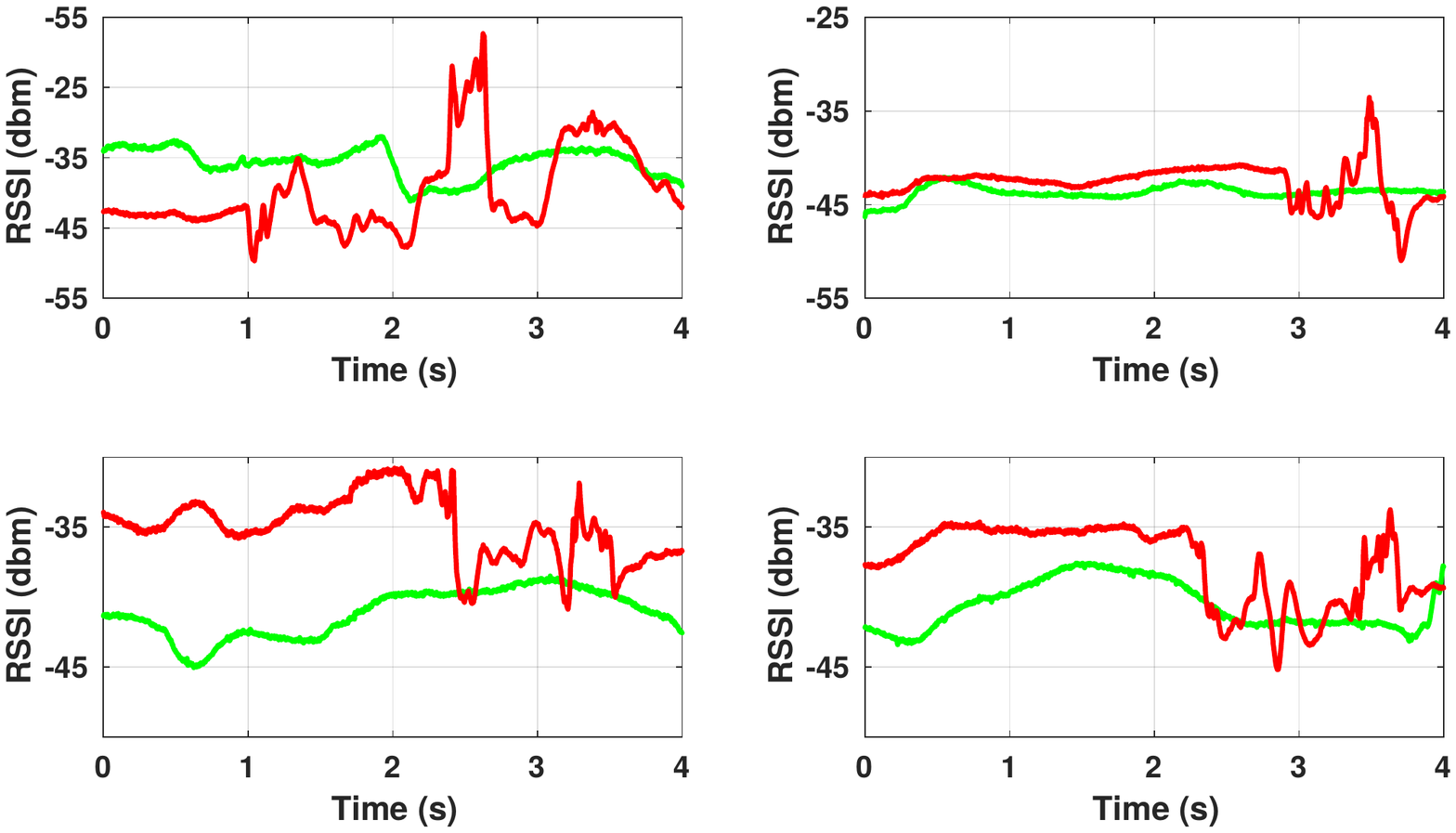}
 \caption{Neutral vs. Angry graphs of subject2}
\end{subfigure}

\caption{Denoised Graphs of driving experiment for neutral case (green) and angry case (red). Each case is driven 4 times by each subject.}
\label{RFgraphs}
\end{figure*}

\begin{figure}
\centering
\begin{subfigure}{.45\columnwidth}
\includegraphics[trim=100 200 0 160,width=0.96\linewidth,height=4cm]{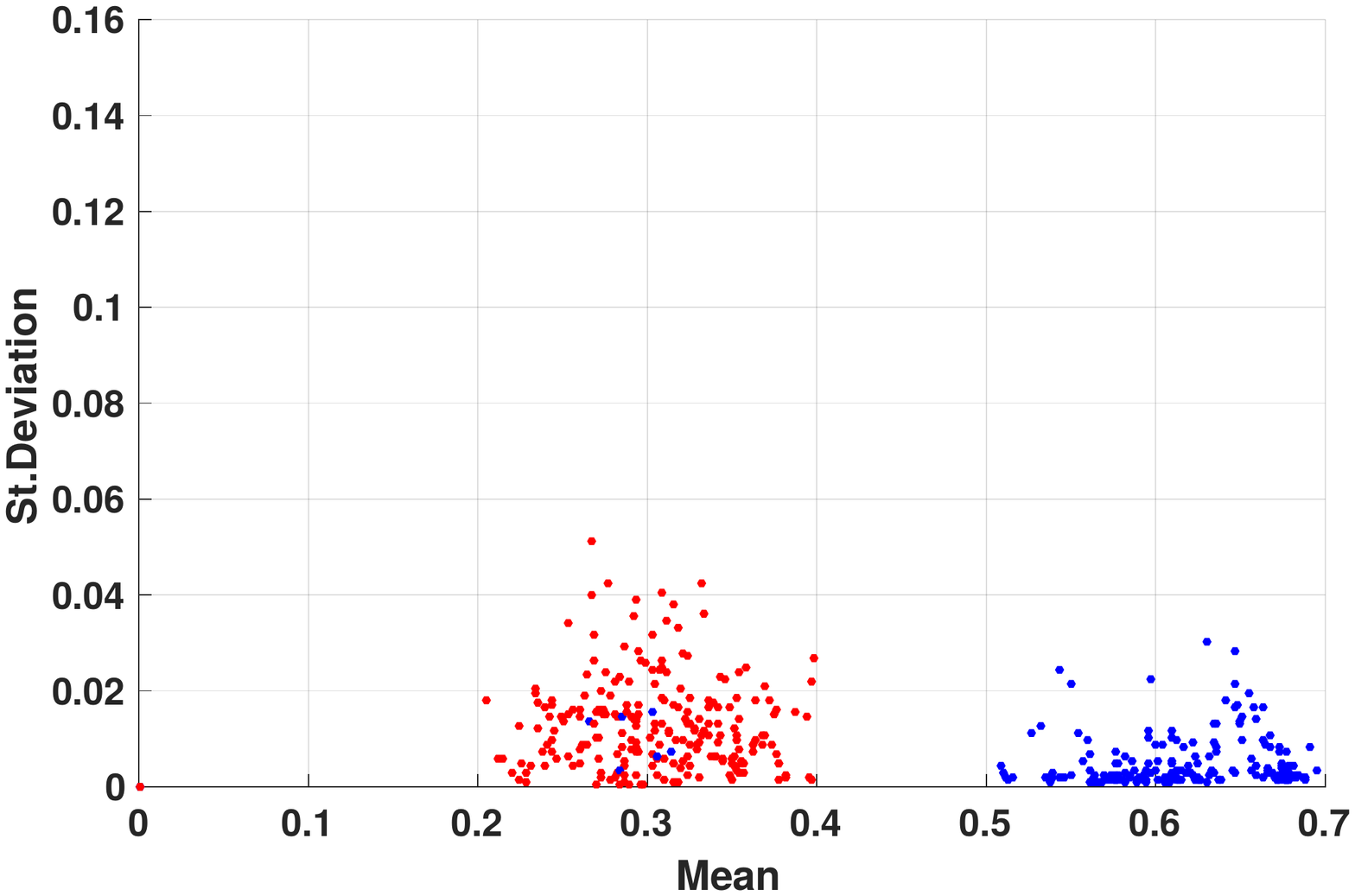}
   \caption{scatterplot of one subject }
\end{subfigure}
\begin{subfigure}{.45\columnwidth}
\includegraphics[trim=40 190 150 150,width=0.96\linewidth,height=4cm]{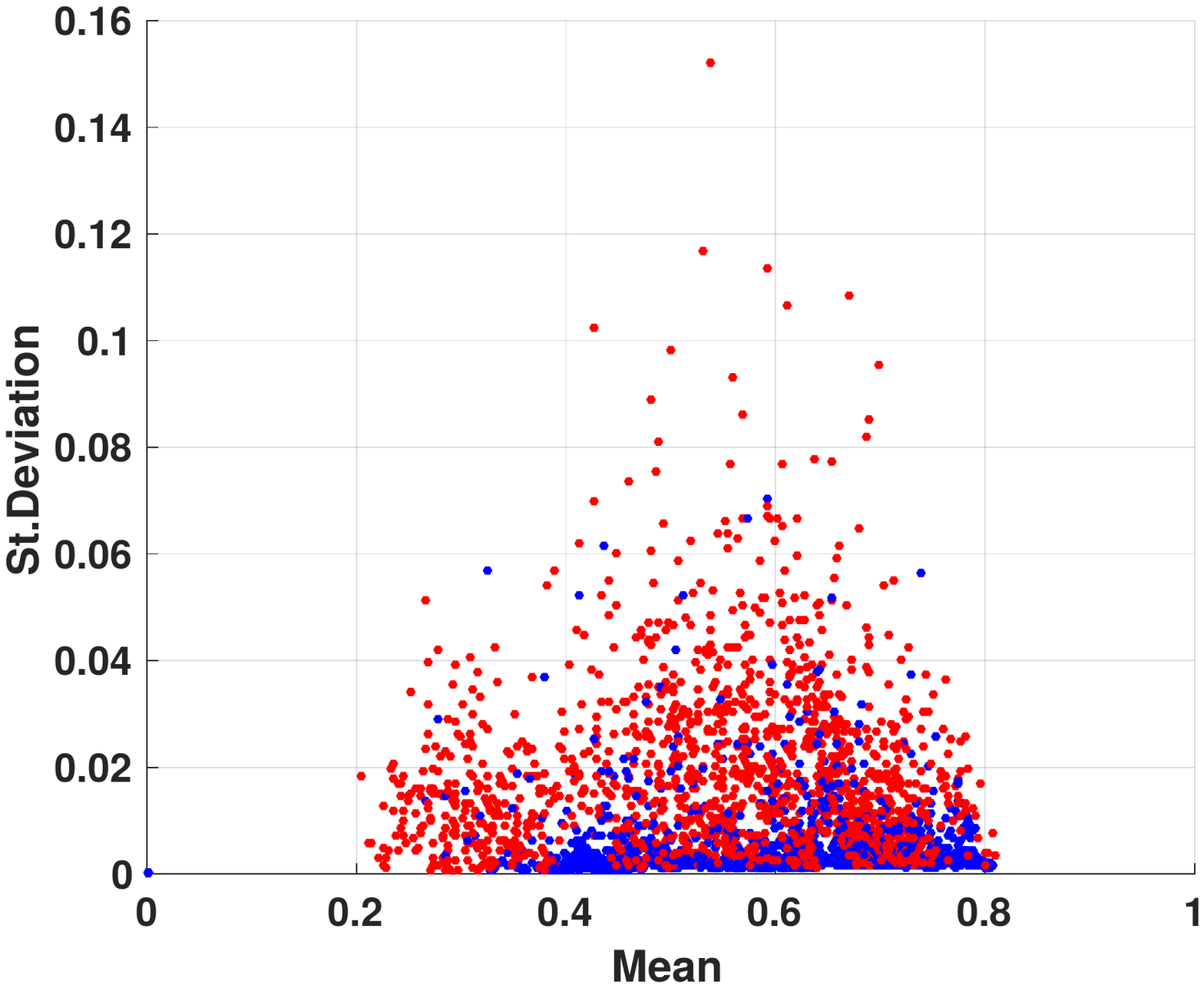}
 \caption{scatterplot of all subjects}
\end{subfigure}

\caption{Scatterplots representing neutral case in blue dots and angry case in red dots. For single subject the plot is highly interleaved leading to accuracy of 98\% while the plot for all subjects overlaps reducing the accuracy to 82.9\%. X and y-axis represent features mean and deviation respectively.}
\label{RFscatter}
\end{figure}

\begin{table}
\centering
\def\arraystretch{2}\setlength\tabcolsep{8pt}
\begin{minipage}{0.5\textwidth}
\centering
\begin{tabular}{|c|r|r|r|} \hline
& $ Neutral $ & $Angry$& $Recall$ \\\hline\hline  
\cellcolor{gray!20}\centering\textbf{Neutral}  & \cellcolor{green!20}97\% & \cellcolor{pink!20}3\%&  \cellcolor{green!10}0.97   \\
 \cellcolor{gray!20} \centering\textbf{Angry} & \cellcolor{pink!20}1\% & \cellcolor{green!20}99\% &  \cellcolor{green!10}0.93 \\
\cellcolor{gray!20} \centering\textbf{Precision}&\cellcolor{green!10}0.94&\cellcolor{green!10}0.96&  \cellcolor{green!10} \\
\hline
\end{tabular}
\caption{Confusion matrix from a single subject's trained model.}
\label{tab:accuracy} 
\end{minipage}%
\hfill
\begin{minipage}{0.5\textwidth}
\centering
\begin{tabular}{|c|r|r|r|}\hline
& $ Neutral $ & $Angry$& $Recall$ \\\hline\hline 
\cellcolor{gray!20}\centering\textbf{Neutral}  & \cellcolor{green!20}85\% & \cellcolor{pink!20}15\%&  \cellcolor{green!10}0.83   \\
 \cellcolor{gray!20} \centering\textbf{Angry} & \cellcolor{pink!20}20\% & \cellcolor{green!20}80\% &  \cellcolor{green!10}0.8\\
 \cellcolor{gray!20} \centering\textbf{Precision}&\cellcolor{green!10}0.84&\cellcolor{green!10}0.78&  \cellcolor{green!10} \\
\hline
\end{tabular}
 \caption{Confusion matrix for all subjects trained model.} 
 \label{tab:ompdiff} 
\end{minipage}
\end{table}

\subsubsection{Results and discussion} 
The graphical representation of processed signals for neutral vs. angry driving is shown in Figure~\ref{RFgraphs} and the scatterplot from classification results are shown in Figure~\ref{RFscatter}. 
The combination of mean and standard deviation gives the best results reaching an overall accuracy of 98\% for individual model and 82.9\% for inter-subject combined data model.
Classification results after 10-fold cross validation are shown in Table~\ref{tab:accuracy} and~\ref{tab:ompdiff}. 
The method is k-NN with 10 neighbours and the distance metric is Euclidean. 
The model preset is Medium Tree with 20 as the maximum number of splits and Gini's diversity index as split criterion.
As we focus to distinguish between the neutral and angry driving behaviour, we see great potential in RF technology for emotion sensing in cars based on the high recognition accuracy. 
Driving attires vary from nationality to nationality and from person to person.
Note, however, that in a vehicular setting, it is reasonable to assume that the recognition system can be trained on the normal driving behaviour over a longer period of time so that inter-subject classification could be an initial default setting only. 

If angry state is being spotted for a very long interval, say 5 to 10 minutes and several times in an hour long journey, then the driver needs to be alarmed and provided with safety guidance. 

Since crossvalidation might be biased when subsequent windows for the feature computation are correlated, we also computed the results with leave-one-subject-out crossvalidation using the same settings as above.
In this case, the overall accuracy achieved for the classification ranges from 71\% to 75\% for the 7 training cases.

\subsection{Scenario 2: Angry Behaviour Detection in a Conversation}
The scenario we consider is an office environment in which a subject carries out a conversation, such as a negotiation, with another subject either on the phone or in person. 
The main subject who is being monitored is standing and free to move during the conversation. 
This environment is more challenging, because it is more flexible than the car environment due to full body movements, diverse surroundings, as well as increased distance between transmitter, subject and receiver. 
\begin{figure}
\begin{subfigure}{.337\columnwidth}
\includegraphics[width=0.9\linewidth]{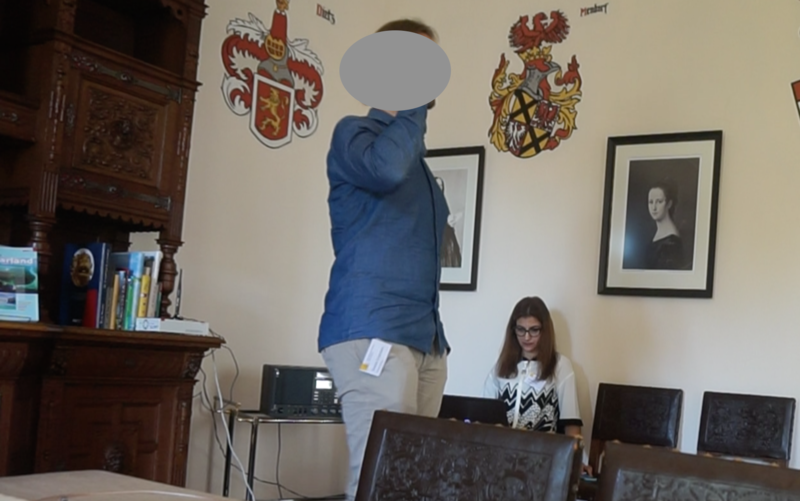}
 \caption{neutral 1 }
\end{subfigure}%
\begin{subfigure}{.337\columnwidth}
\includegraphics[width=0.9\linewidth]{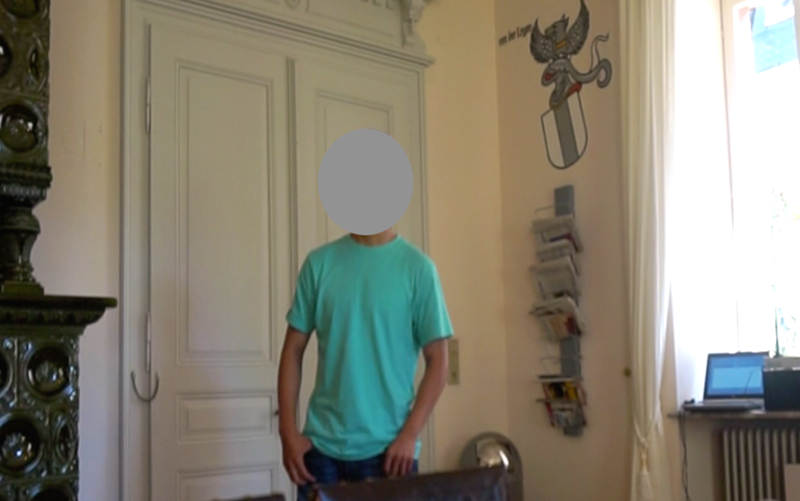}
\caption{neutral 2}
\end{subfigure}%
\begin{subfigure}{.337\columnwidth}
\includegraphics[width=0.9\linewidth]{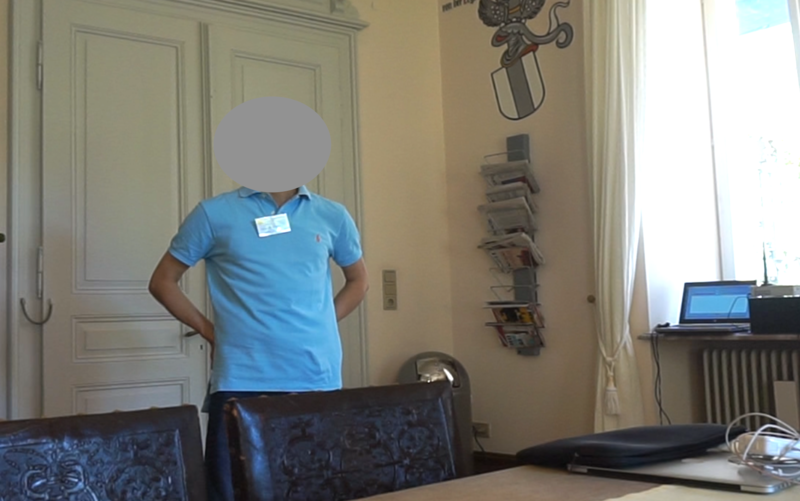}
\caption{neutral 3}
\end{subfigure}%

\begin{subfigure}{.337\columnwidth}
\includegraphics[width=0.9\linewidth]{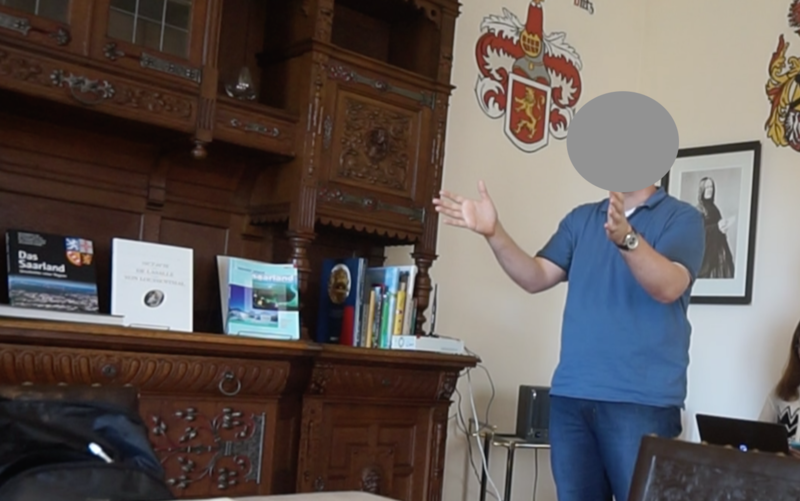}
\caption{angry 1}
\end{subfigure}%
\begin{subfigure}{.337\columnwidth}
\includegraphics[width=0.9\linewidth]{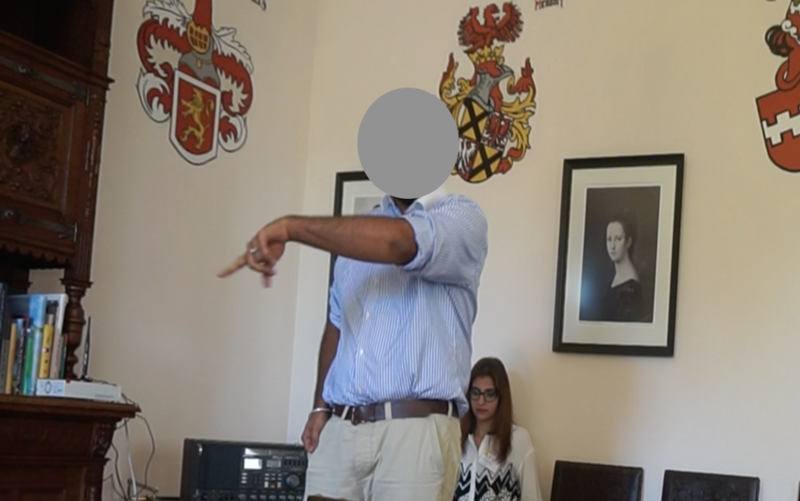}
\caption{angry 2}
\end{subfigure}%
\begin{subfigure}{.337\columnwidth}
\includegraphics[width=0.9\linewidth]{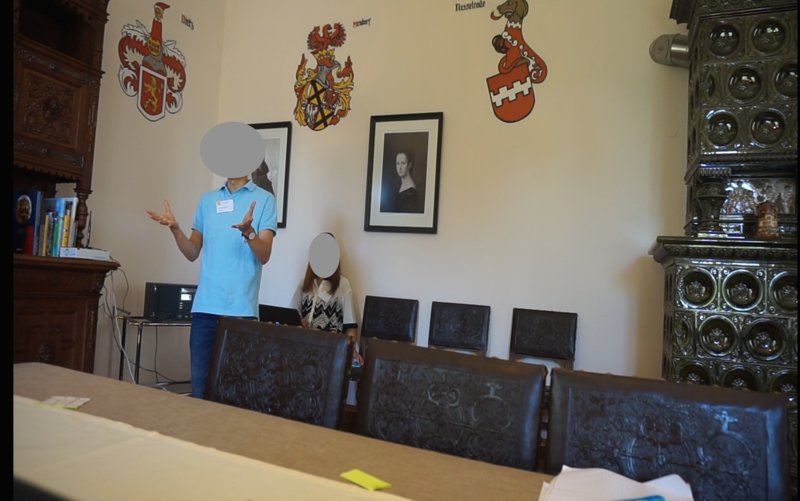}
\caption{angry 3}
\end{subfigure}%
\caption{Images captured during the conversation experiments, focussing on the main subject. The top row indicates the body gestures in a neutral state during a conversation. They are more static and controlled. While the bottom row indicates the random gestures during conversation in angry states. The gestures are lifted, strong and more frequently performed.}
\label{dagImages}
\end{figure}
\subsubsection{Experiment}
5 subjects, different from previous experiments in Section \ref{snrSection} and \ref{carSection} participated in the experiment. 
Participants nationalities were German, China and India. 
Images from the experiment are shown in Figure~\ref{dagImages}. 
We intentionally change the distance between receiver and subject for more general recognition case, keeping the SNR value and distance between receiver and transmitter constant throughout the experimentation. 
The space is a large meeting or conference room of about 12x18 sqft with wooden furniture tables, chairs, computers etc. 
The distance between the transmitter and receiver is 8m, SNR is 42dB and all the equipment, configuration, data collection and pre-processing are the same as for the driving experiment in Section~\ref{carSection}. 
For each subject, the first neutral vs. angry data is captured by keeping 2m distance between receiver and subject. 
The measurements are taken twice for each emotional state. 
Then the distance is increased to 5m between receiver and subject, and the same experiment is performed. 
The whole conversation is captured on video for validation and ground-truth labelling. 
Each subject is instructed separately and free to decide on how they want to do a conversation. 
Two of the participants chose to talk on the phone and 3 of them preferred talking in person with another subject. 
They were also asked about the situations that make them irritable and angry. 
As all the subjects belong to the research domain, their situations were relevant to research publications, colleagues and students they deal with. 
Our subjects came up with a few scenarios where they would feel anger. 
Such anger inducing scenarios included \textit{not being given due credit for hard work}, or \textit{people not keeping time for meetings or being late without any acceptable reason}. 
Based on the individual discussions, we constructed a cover story for each one of them to best induce the anger emotion. 
An example cover story for each neutral and angry state is described below:

\vspace{.3cm}
\noindent\fbox{\begin{minipage}{.95\columnwidth}
\textbf{Neutral Case:}  \textit{``Your friend is waiting at the main corridor of your office building and cannot find a way to your room, give him suitable instructions to help him navigate easily."}\\
\textbf{Angry Case:}  \textit{``You have been working on a project with your coworker for a year. Your coworker secretly publishes the work without giving you the credit in the article. You discover this article online and this news makes you furious. You call him to your office and have an argument with him on this matter."}
      \end{minipage}}\vspace{.3cm}

\begin{figure*}
\centering
\begin{subfigure}{.45\textwidth}
  \centering
  \includegraphics[trim=20 200 0 255,width=1.13\linewidth,height=5.5cm]{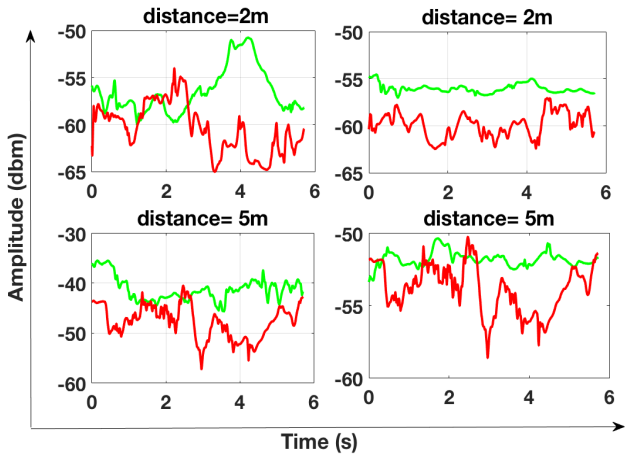}
   \caption{Neutral vs. Angry graphs of subject1 }
\end{subfigure}
\begin{subfigure}{.45\textwidth}
  \centering
  \includegraphics[trim=0 200 0 255,width=1.16\linewidth,height=5.5cm]{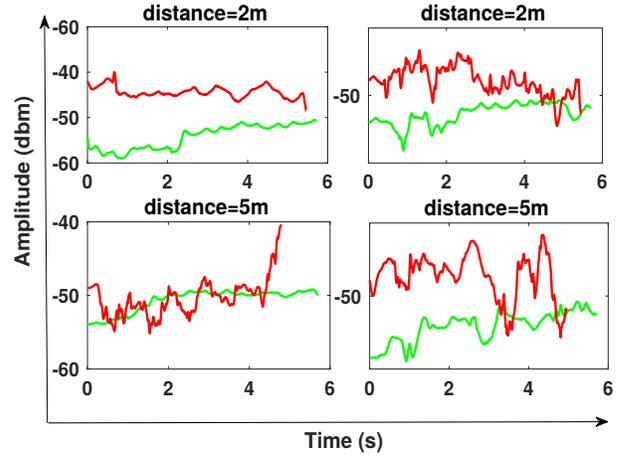}
 \caption{Neutral vs. Angry graphs of subject2}
\end{subfigure}
\caption{Denoised Graphs of conversation experiment for neutral case (green) and angry case (red). Each case is repeated 4 times by each subject.}
\label{dagPlots}
\end{figure*}

\begin{table}
\def\arraystretch{2}\setlength\tabcolsep{3.5pt}
\begin{minipage}{0.5\textwidth}
\begin{tabular}{|c|r|r|r|r|c|r|r|r|r|}\hline

\rowcolor{lightgray}&\multicolumn{3}{c|}{Distance= 2m} &&\multicolumn{3}{c|}{Distance= 5m} \\
& $ Neutral $ & $Angry$ &$Recall$&&$Neutral$& $Angry$ & $Recall$ \\\hline\hline 
\cellcolor{gray!20}\centering\textbf{Neutral}  & \cellcolor{green!20}89\% & \cellcolor{pink!20}11\%&\cellcolor{green!10}0.86 &\cellcolor{gray!20} &\cellcolor{green!20}96\%&\cellcolor{pink!20}4\%&\cellcolor{green!10}0.83 \\
 \cellcolor{gray!20} \centering\textbf{Angry} & \cellcolor{pink!20}22\% & \cellcolor{green!20}78\% & \cellcolor{green!10}0.83 &\cellcolor{gray!20} &\cellcolor{pink!20}29\%&\cellcolor{green!20}71\%&\cellcolor{green!10}0.79 \\
 \cellcolor{gray!20}\centering\textbf{Precision} & \cellcolor{green!10}0.9& \cellcolor{green!10}0.76&\cellcolor{green!10}&\cellcolor{gray!20} & \cellcolor{green!10}0.8&\cellcolor{green!10}0.82&\cellcolor{green!10}
 \\
\hline
\end{tabular}
\end{minipage}
\captionof{table}{Confusion matrices from a single subject's trained model. Left: 2m distance between between receiver and subject; Right: 5m distance.}
\label{accuracy2} 
\end{table}

  \subsubsection{Results and Discussion}
\begin{figure}
\centering
\begin{subfigure}{.45\columnwidth}
\includegraphics[width=0.98\linewidth]{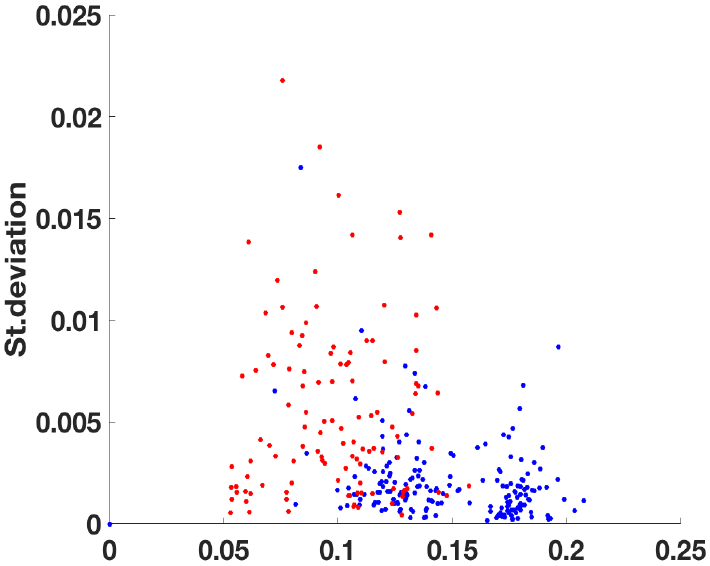}
   \caption{distance= 2m }
\end{subfigure}
\begin{subfigure}{.45\columnwidth}
\includegraphics[width=0.98\linewidth]{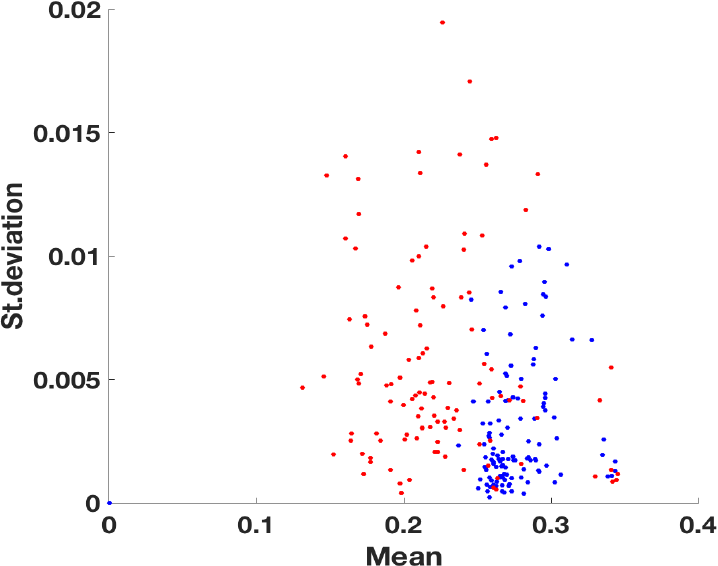}
 \caption{distance= 5m }
\end{subfigure}
\caption{Scatterplots representing neutral case in blue dots and angry case in red dots. The left plot is the result when the distance between receiver and subject is 2m. Second plot represents the data results when the distance between receiver and subject is 5m. The accuracy is not affected by increasing 3m distance in this case.}
\label{scatterPlot2}
\end{figure}

The results of the classification are shown in Table~\ref{accuracy2}, for both 2m distance and 5m distance between receiver and subject. 
As the SNR value is high enough for gesture recognition (as described in Table \ref{accTable}), even at 5m distance, it is clear that the body gestures are promising indicators of anger. 
The overall accuracy has, however, decreased (maximum for individuals 84.9\%) as compared to individual results for driving experiments. 
The reason is that car is a closed space and the driver does not perform any significant lower body movements. 
However, in this case the subject can move around and can do whole body movement within a range of specified distance between receiver and subject. 
The graphs in Figure~\ref{dagPlots} represent the difference of signals tracked between neutral case and angry case for two subjects. 
For each subject, the first part of the graph shows results when distance between subject and receiver is 2m. 
The second part shows the results for 5m distance between receiver and subject. 
Scatterplots for one subject in Figure~\ref{scatterPlot2} also show slight overlapping between angry and neutral case. 
Inter-subject classification for all subjects reduces the accuracy down to 64\% for this scenario. 

\section{Conclusion}\label{sectionConclusion}
We have presented \textit{RFexpress!}, the first-ever network edge based Device-free motion and gesture-based emotion sensing system. 
The system has been exploited in a two scenarios where wireless access points (such as WIFI, bluetooth, Wlan APs etc.) are now a days fundamentally deployed at large scale. These scenarios are vehicular edge scenario and stationary edge scenario; to detect risky driving behaviour and to distinguish between neutral and angry human communication respectively. 
The experiment on the risky driving detection is performed by 8 drivers and each emotional state is experimented 4 times for each driver. 
The angry emotion is induced using well established emotion elicitation techniques. 
We observed that a car has well suited conditions for DFAR-based emotion recognition: high SNR, fairly constant environment and limited number of people. 
RF signals processing at these edge devices make emotion sensing from body gestures( via Wifi signals) very promising for vehicular settings.  
We could achieve an accuracy of 98\% in the vehicular setting. 
In the second, office scenario, we considered the distinction of angry versus neutral conversation in non-scripted realistic environments with increasing distance between subject and receiver. 
In this experiment we could achieve an accuracy of up to 82.9\% for individually trained models and 64\% for inter-subject models. 

In addition, we studied the network edge characteristics influencing DFAR, in particular gesture and emotion recognition, in real environments as compared to controlled environments. 
We measure the radio characteristics in different environments like cafe, outdoor, malls and office space. 
Then we model these radio characteristics in our lab and perform gesture recognition experiment to analyse the variation in its accuracy with changing modelled radio characteristics. 
With SNR as our primary indicator, we consider 6 different SNR values, (59dB, 42dB, 22dB, 12dB, 2dB and 0dB).  
In order to model the required SNR values, we use a USRP based transmitter and receiver se-tup. 
The power values of the transmitted signal are measured using a spectrum analyser at the receiver end.  
The SNR values are obtained by configuring power and noise values at a transmitter using GNU radio. 
Three gestures, hands down, hands raised and clapping are detected at each SNR value. 
This experiment is performed by 5 participants 5 times each for every activity. 
The distance between the transmitter and receiver (2m) and all the other parameters are kept constant throughout the experimentation. 
The results show that accuracy above 80\% can be achieved at SNR higher than 30dB. 
At SNR 20db and below, the accuracy of gesture recognition drops significantly. 
This accuracy also varies with the complexity of gesture performed. 
For a simple hands down and hands up case, accuracies are fairly high. 
More complex activities, however, like clapping, can only be detected at 40dB or higher. 
We also double the distance between transmitter and receiver and perform gesture recognition. 
The accuracy is no affected by increasing distance between transmitter and receiver up to 8m, as the SNR remains considerably high (~59db) within 10m. 

\section{Future work}\label{sectionFuturework}
In the future we can scale our  \textit{RFexpress!} solution and deploy it as a service in the network edge, where it can benefit from the computational resources of the edge to perform more complex DFAR operations. In this scenario, receiver devices with less processing power such as low end WIFI access points or mobile phones can upload their received signal data to the  \textit{RFexpress!} service in the edge which would then process the signal, extract features and derive emotion information from the data. This information could be then fed back to the users in real time.
Also this opens up possibilities where multiple receiver devices such as two WIFI access points in different ends of a room, can cooperate with each other by sharing their collected data and thus improving the overall accuracy of the system. 

Moreover, we intend to further explore the emotion sensing with RF technology. We extend our detected emotions from neutral vs. angry to neutral, angry, happy, tired and sad.  
The immediate aim is to achieve higher accuracy for our car driving scenario for more subjects with varying demographic profiles. Another critical human emotion or behaviour is tiredness, in a car driving scenario. We would detect the tired state and also differentiate between less angry and extremely angry behaviour in order to generate an appropriate feedback.

\section*{Acknowledgment}
The authors would like to acknowledge the valuable contribution, support and guidance by Syed Safi Ali Shah in this work. 
Our sincere gratitude to all the participants who greatly helped in carrying out experiments. 

\bibliographystyle{IEEEtran}


\end{document}